# Presenting the Sense of Effort through Vibration Based on Force Estimated by Inverse Dynamics in Videos

Ryoma Akai[1], Masashi Konyo[1], and Satoshi Tadokoro[1]

[1] *Graduate School of Information Sciences, Tohoku University, Japan*

(Email: konyo@rm.is.tohoku.ac.jp)

**Abstract ---** We present the sense of effort through vibration to help the video viewer understand how the person in the video moves the body. We suppose sense of effort is related to force, so we generate vibration based on force and present the sense of effort through the vibration. We use perceived intensity to make sense of effort proportional to vibration. In our demonstration, you can experience vibration while watching a video. We can create vibration on the spot, so you can experience vibration made from a video taken on the spot.

## 1 INTRODUCTION

In recent years, video has become a mainstream media due to the widespread use of smartphones and the development of video distribution platforms. Anyone can easily become a video creator, and a wide variety of videos can be posted on video distribution platforms.

People use videos in situations where they learn how to move their bodies, such as in sports. For example, in the case of dance, video materials of dance and videos of professional dancers are posted. Thus, beginners can learn choreography by watching videos of skilled dancers. However, it is difficult for beginners to notice subtle differences in the movements just by watching the videos. Because dance is a complex movement, and the whole body moves, it is challenging to follow the movements of specific body parts, such as the torso, through visual observation alone.

Therefore, to make human actions more straightforward to understand, there is a need for technology that adds information about human actions other than visual information. The physical load felt during exercise is influenced by moving joints and gravity. The perception of force and the perception of effort are two perceptions closely linked to another [1]. In this paper, we focus on the force acting on the joints and aim to convey the sense of effort using vibration based on the force.

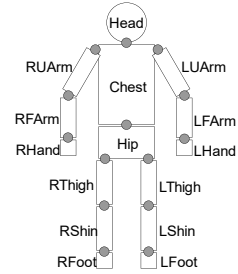

Fig.1  Link model to calculate inverse dynamics [3].

Table 1  Mass of each body part [3].

| Body part | Mass ratio | Body part | Mass ratio |
|---|---|---|---|
| Head | 0.069 | Hip | 0.187 |
| Chest | 0.302 | RThigh | 0.110 |
| RUArm | 0.027 | RShin | 0.051 |
| RFArm | 0.016 | RFoot | 0.011 |
| RHand | 0.006 | LThigh | 0.110 |
| LUArm | 0.027 | LShin | 0.051 |
| LFArm | 0.016 | LFoot | 0.011 |
| LHand | 0.006 | | |

## 2 METHOD

We propose a method for generating vibration that match the sense of effort of a person in any video. The flow is extracting motion from the video, estimating the force, supposing sense of effort, relating to perceived intensity, and generating vibration based on the intensity.

We extracted motion from video by BlazePose [2], a human pose estimation library. It can obtain the position of the skeleton. To estimate the force acting on the joints using inverse dynamics, we used the link model from [3] shown in Fig. 1. Table. 1 shows the mass ratio of each part [3]. Using the internal division ratio [4] shown in Fig. 2, we obtained the position of the center of gravity of each part of the link model from the position obtained by pose estimation. The red dots in Fig. 2 are the positions obtained by BlazePose, and the blue dots are the center of gravity positions of each part of the link model. We calculated the acceleration of the center of gravity from the position of it. The acceleration was smoothed and differentiated using a Savitzky-Golay filter [5]. The equations of motion for translation were used to estimate the force acting on the joints moving each site: the equation of motion for translation of site $i$ connected to sites $i-1$ and $i+1$ at the two joints is Equation. 1.

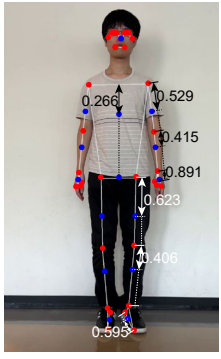

Fig.2 The relationship between the center of gravity of each part in the link model and landmark estimated by BlazePose.

$$m_i \ddot{x}_{gi} = m_i g + f_i - f_{i+1} \quad (1)$$

where $m_i$ is the mass of site $i$, $x_{gi}$ is the position vector of the center of gravity of site $i$, $g$ is the gravity acceleration vector, $f_i$ is the force acting on the joint connecting site $i-1$ and site $i$.

If no force is acting on the terminal joint, $f_{i+1}$ becomes 0, and we can obtain the force $f_i$. By solving sequentially, we can obtain the force acting on each joint.

The flow of vibration waveform generation is as follows: to capture the magnitude of the motion, we took the absolute value of the force and normalized it. According to Stevens' power law [6], which describes the relationship between the magnitude of a physical stimulus and its actual perceived intensity, the exponent of power corresponding to force is 1.7. Therefore, we set the sense of effort to the 1.7 power of the force. We proportionated perceived intensity of vibration to sense of effort. The AM wave presented was the sum of the intensity per segment of the original signal converted to amplitude at a frequency of 200 Hz from Equation. 2 [7].

$$A(f) = A_T(f) I_{total}^{\frac{1}{2\alpha(f)}} \quad (2)$$

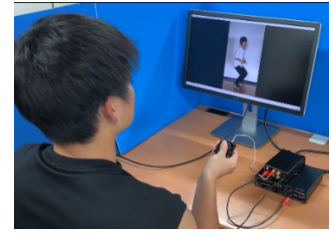

Fig.3 The appearance of demonstration

where $A(f)$ is the amplitude of the vibration at frequency $f$, $A_T(f)$ is the amplitude of the vibration detection threshold at $f$, $I_{total}$ is the sum of the intensity per segment of the original signal, and $\alpha(f)$ is an exponential parameter representing frequency dependence.

Fig. 3 shows the appearance of demonstration. You can feel vibration with a hand-held vibrator.

## 3 CONCLUSION

In this study, we proposed a method for generating vibration from the force generated by motion to present sense of effort of a person in a video. In our demonstration, you can experience vibration generated based on the ground reaction force of the squat. So, you can feel the sense of effort of the centroid.